\documentclass{aa}
\usepackage[dvips]{graphics}
\usepackage{supertabular}

\begin{document}
\newcommand{\ch}[1]{\underline{#1}}
\newcommand{\cthead}[1]{\multicolumn{1}{c}{#1}}

\title{Investigation of flat spectrum radio sources by the interplanetary
scintillation method at 111~MHz.}
\author{S.A.Tyul'bashev$^{1,2}$ \and P.Augusto$^3$}
\offprints{S.A.Tyul'bashev} \institute{ 1) Pushchino Radio
Astronomy Observatory, Pushchino, Moscow region, 142290 Russia\\
2) Isaac Newton Institute, Chile, Pushchino Branch, Russia\\
3)Universidade da Madeira, Centro de Ci\^encias Matem\'aticas,
Caminho da Penteada, 9000-390 Funchal, Portugal}

\date{}
\maketitle

\abstract{Interplanetary scintillation observations of 48 of the
55 Augusto et al. (1998) flat spectrum radio sources were
carried out at 111~MHz using the interplanetary scintillation
method on the Large Phased Array (LPA) in Russia. Due to the
large size of the LPA beam ($1\degr \times 0.5\degr$) a careful
inspection of all possible confusion sources was made using extant
large radio surveys: 37 of the 48 sources are not confused. We were able to estimate the
scintillating flux densities of 13 sources, getting upper limits
for the remaining 35. Gathering more or improving extant VLBI
data on these sources might significantly improve our results.
This proof-of-concept project tells us that compact ($<1''$)
flat spectrum radio sources show strong enough scintillations at
111~MHz to establish/constrain their spectra (low-frequency end).

}
\authorrunning{S.A.Tyul'bashev \& P.Augusto}
\titlerunning{ Flat spectrum radio sources at 111~MHz.}


\section{Introduction}

A systematic search for dominant structure on
0.09--0.3$''$ scales in large flat-spectrum radio source samples
was made by Augusto et al. (\cite{Augusto98}). Fifty-five radio
sources were selected from a parent sample containing 1665 strong
flat-spectrum radio sources ($S_{8.4~GHz}>100$~mJy;
$\alpha_{1.4}^{4.85}<0.5, S_{\nu}\propto\nu^{-\alpha}$). These
sources all have published MERLIN 5~GHz data. A few also have VLBA
5~GHz and MERLIN 22 GHz maps (Augusto et al. \cite{Augusto98}). In
addition, some others have MERLIN+EVN 1.6 GHz high angular
resolution ($<0.5''$) unpublished data (Augusto et al., in
prep.).

The study of these 55 sources is not complete without low
frequency observations ($\sim100$ MHz), as was pointed out in
Augusto et al. (\cite{Augusto98}), where the spectra of most
sources have no data at all below $\sim300$ MHz. The turnovers in
the spectra of compact components in these sources must be found,
to give a physical meaning to all 55 sources, namely by fitting
synchrotron emission spectra for them all. Since VLBI does not
routinely (or efficiently) operate at such low frequencies, we use
the interplanetary scintillation (IPS) method at 111 MHz with the
Large Phased Array (LPA). Very similar work was done at LPA for
compact steep spectrum sources (Artyukh et al.\ \cite{Artyukh99};
Tyul'bashev \& Chernikov \cite{Tyulbashev00},
\cite{Tyulbashev01}). The principle of IPS is very simple:
the solar wind has variations in electron density on which
depends the velocity of the radio waves that travel through it.
As a result, we have a phase screen which can increase or decrease
the signal from distant radio sources; i.e. the sources will {\em
scintillate}. The characteristic time of scintillations depends
on the velocity of the solar wind, on the frequency of the
observations, and on the sizes of the electron clouds. For example,
if we have observations at 111~MHz, this characteristic time scale
is approximately one second. The scintillations will be stronger
if the distant radio sources (or components therein) have small
angular sizes ($<1''$). Details of observations by the IPS method
and relevant theory can be found, for example, in Vlasov et al.\
(\cite{Vlasov79}).

The IPS method has advantages and disadvantages when compared with
VLBI observations. The main advantage is the possibility to
observe sources at low frequencies and high resolution. The
main disadvantage is the very low positional accuracy. We see
scintillations, but we do not know exactly which component(s)
is(are) scintillating or even if we correctly identify the main
radio source (among many in-beam): the coordinate
uncertainties for LPA are 5-10$^s$ in right ascension and
2-3$^\prime$ in declination for strong sources
($\sigma_{scint}/\sigma_{noise} > 2$, at $\tau=0.5$~s; standard
SNR $>7$), increasing to 30-60$^s$ and 5-7$^\prime$,
respectively, for weak sources
($\sigma_{scint}/\sigma_{noise} < 2$). These uncertainties
have a complex behaviour ($f(\sigma_{scint},\sigma_{noise}$);
Artyukh \& Tyul'bashev \cite{Artyukh96}). In order to get around
these large, inherent errors (i.e. to, at least, correctly identify
the source with the scintillating component) for all
pointings that we have done with the LPA, we extensively searched
the entire beam area using both the 1.4~GHz NRAO VLA Sky Survey
(NVSS; Condon et al.\ \cite{Condon98}; www.cv.nrao.edu/nvss) and
the 74~MHz VLA Low-Frequency Sky Survey (VLSS;
lwa.nrl.navy.mil/VLSS). The whole of Sect.~3 is devoted to this
study. In Sect.~2, we present both the data collection and
reduction, while in Sect.~4 we compile the IPS 111 MHz results
from observations of 48 of the 55 Augusto et al.\
(1998) sources, for which we derive either a scintillating
flux density estimate (13) or an upper limit (35). We also
include in this section, as a case study, the detailed spectrum
analysis for B0821+394. Finally, a short discussion and
summary is given in Sect.~5.

\section{Observations and data analysis}

We carried out 111~MHz IPS observations with the LPA (a meridian
instrument) of the Lebedev Institute of Physics, Russia. The
effective area\footnote{ Due to the large number of parameters
on which the effective area depends, it can actually change by up
to 20--30\% from day to day.} of the antenna in the zenith
direction is $2 \times 10^{4}$ m$^{2}$ with a beam
approximately $1\degr \times 0.5\degr$ (EW $\times$ NS) in
size. The receiver integration time was $\tau=0.5$~s, the
sampling time 0.1~s, and its bandwidth 600~kHz. As a result,
the sensitivity of LPA for scintillating sources is
$\sigma_{scint}\simeq$0.15--0.2~Jy in the zenith direction
(with s.n.r. $\ge 10$, after the integration of all
scintillations\footnote{Although we start up with
$\sigma_{scint}=\sigma_{noise}\simeq0.2$ and $\tau=0.5$,
the 9--18 mins integration times assure 1080--2160 independent points.
Since the s.n.r increases with the square root of these, we get
s.n.r $\simeq30$--45 which, being conservative, we translate into
s.n.r. $\ge 10$.}), decreasing with source declination
as $\cos(\delta)$, where $\delta$ is the declination of the source.
The r.m.s. confusion due to extended (nonscintillating) sources is
$\sim$1 Jy, while the r.m.s. confusion due to scintillating
sources is $\le$ 0.12~Jy. This means that even when it is
difficult to measure the total flux density of a source, it is
still possible to measure the scintillating flux density.

We carried out 137 sessions in 2001-2002, each with a duration
between 5 and 11 hours. We observed, in each session, from 5 to 10
calibrators\footnote{We have amplitude calibrated the observations
using many radio sources from the 3C/4C catalogues. All
flux-density estimates were made in the scale of Kellermann
(\cite{Kellermann64}).} and always less than 15 target sources.
Thus, a total of between 20 and 25 individual records were
gathered per session and all targets were observed in more
than one run ($N$ on Column~(2) of Table~3). The integration time
for each source depended on its declinitaion, so it varied from
approximately 9 to 18 minutes. In total, we had over 1100~hrs of
observation time, half of which on-target. Many individual source
observations had to be prolonged to compensate for interference.
Due to the large number of sources, it was not possible to
choose the best elongation for each source as it was done in
Artyukh (\cite{Artyukh81}). Therefore, we used the converting
coefficients of Marians (\cite{Marians75}). We also selected the
best data: the records, among the many observed, with the lowest
noise.

Flat spectrum sources are very difficult to detect at low
frequencies (in total flux density), therefore the data
reduction must be made with care (c.f. similar steep-spectrum
radio source analysis in Tyul'bashev\& Chernikov
\cite{Tyulbashev01}). The data reduction method we
used is given in Artyukh (\cite{Artyukh81}) and Artyukh \&
Tyul'bashev (\cite{Artyukh96}). This method enables us to detect
faint scintillating sources, for which the scintillation
dispersion ($\sigma_{scint}^2$) is smaller than the noise
(dispersion) on the receiver time constant $\tau$. We
estimate this noise in the parts of the data record where we
cannot see scintillations, i.e., where the noise seen is minimal.
The accuracy of the scintillating flux density estimate
($S_{compact} \equiv S_c$) depends on the fluctuation of the
flux density ($ \sigma_{scint}$) and on the elongation of
the source (angle between the Sun and source directions as seen by
the observer). The typical accuracy is 20-25\% for elongations
smaller than $40\degr$ and ($ \sigma_{scint}$) higher than
the noise of the antenna in a given direction. In the worst
cases, the accuracy of $S_c$ estimates is still better than
30-50\% (see details in Artyukh \& Tyul'bashev \cite{Artyukh96};
Artyukh et al. \cite{Artyukh98}).


Our observations lead to two situations: i) the compact
source/components is/are too weak; no scintillations are detected
but we can place an upper limit on the scintillation flux density
($S_c$); ii) scintillations are seen from a compact component in
the source. We try to get the best possible estimate of $S_c$ by
combining all existing (good) records (column (2) of Table~3). The
individual (statistical) error of a single record is 5--7\%, hence
combining them decreases it. Unfortunately, this error is
overwhelmed by the calibration error\footnote{ There is a
third, nastier error due to bursts from the Sun which can only be
overcome by averaging many records.} at LPA (10--20\%).

In what follows we summarize the observing/data reduction
steps for each source (see also Sect.~4.2):
\begin{enumerate}
\item We observe one (or more) flux density calibrator(s) --- several records.
\item We observe the target source (several records).
\item If possible, we estimate the total flux density ($S_t$) using the calibrator and target records. $S_t$ adds the scintillating flux density (compact component(s)) and the non-scintillating one (extended component(s)).
\item We look for scintillations in the target record by first removing the background and then pulse interferences, having only noise left (instrumental --- $\sigma_{noise}$ --- and scintillating --- $\sigma_{scint}$). Then, we split these noises from the fact that $\sigma_{noise}$ exists all the time, while $\sigma_{scint}$ exists only from a given direction --- {\em primary} record.
\item It is this latter part (few minutes) of the main record that is used to estimate $S_c$ using $\sigma_{scint}$ and information on the angular sizes of the source and its components (e.g. Marians \cite{Marians75}).
\end{enumerate}

\section{Confusion analysis}

The fact that the LPA has a huge beam
($1.0\degr\times0.5\degr$) makes it imperative that we clearly
identify the source that includes the component actually
scintillating at 111~MHz. It might not be the main source (as
listed in Table~3) since many other radio sources exist inside the
LPA beam and might cause confusion due to producing stronger
scintillations. Ideally, we should have available VLBI maps for
all compact (and fairly strong) sources inside each of the LPA
pointings. There is no such survey available at high frequencies
and even fewer at 111~MHz. Hence, the best we can do is to use
existing literature and non-VLBI survey information in order to
guess the source where the scintillating component lies. The best
surveys to date that could suit our purposes used the VLA-A
at 8.4~GHz (0.2\arcsec resolution): the Jodrell-VLA Astrometric
Survey (JVAS; e.g. Patnaik et al.\ \cite{Patnaik92}) and the
Cosmic Lens All-Sky Survey (CLASS; e.g. Myers et al.\
\cite{Myers03}). Apart from the main sources, which all have VLA-A
8.4~GHz compact components, only four ``candidates''\footnote{
In the context of this Section, a {\em candidate} is a source,
inside each LPA pointing, that competes with our main source for
the scintillations that we have observed (Table~1 vs.\ Table~3).}
were detected by those surveys (see below).

There are three surveys of interest to our study. Although
with much lower resolution than JVAS/CLASS, they were made at
lower frequencies. The most relevant of these, at least as regards
the frequency of observation, is the VLSS done with the VLA (B
and BnA) at 74~MHz (80\arcsec resolution). It certainly can
identify the strongest sources in each of our LPA pointings but,
unfortunately, it cannot tell us much about compactness. Another
useful survey is the Faint Images of the Radio Sky at
Twenty-centimeters (FIRST --- Becker, White \& Helfand
\cite{Becker95}; sundog.stsci.edu/top.html), done with the VLA-B
at 1.4~GHz (5\arcsec resolution). Its resolution, altough still
three orders of magnitude above VLBI scales, is 16 times better
than the one of VLSS, but the shift to high frequencies does not
help much in our study. Unfortunately, both of the previous
surveys lack full-sky coverage. The VLSS is still on-going, while
FIRST covers less than half of the northern sky, where all our
sources lie. As a result, out of the 48 pointings done with the
LPA (centred on each of the main sources), 34 (71\%) fell inside
the VLSS sky coverage while only 19 (40\%) are in FIRST. The last
survey we used in our study is the NVSS, made with the VLA (D and
DnC) at 1.4~GHz (45\arcsec\, resolution), which covers the full
northern hemisphere; hence, it should contain {\em all} candidates
to confusing sources of our observations.

Our ``candidate-finding'' scheme was to fully examine a
$1.0\degr\times0.5\degr$ area (equal to the LPA beam), centred on
our main source position, using the Internet search engines in
VLSS, NVSS, FIRST, and NED (the NASA Extragalactic Database;
nedwww.ipac.caltech.edu), in order to get extra literature
information (namely radio spectra and maps), if any. This has
found a total of 1046 candidates for the 48 sources or
`pointings' \footnote{ In this Section we use the word
`pointing' to refer to each beam area to be analysed: each
$1.0\degr\times0.5\degr$ area centred on each main source of the
48 observed and listed in Table~3.} ,the vast majority quite weak
(Appendix~A). All of these are in the NVSS, but only 271 (out of
the surveyed total of 378 --- 72\%) and 29 (total 736, so 4\%) are
{\em detected} in FIRST (S$_{1.4}\ga1$~mJy/beam) and VLSS
(S$_{74}\ga0.5$~Jy/beam), respectively. A total of 135 candidates
are in both surveyed areas, bringing the grand total of candidates
with more information than NVSS-only to 979, so only 67 (7\%) lack
it. Three candidates have only non-FIRST maps available while 36
others have only radio spectra as extra information: there are 82
candidates with spectral information of which 27 also have radio
maps (see Table~1). The question now is: {\it How do we know if a
candidate is strongly scintillating or not at 111~MHz?} Obviously,
the seven sources of Table~1 with high resolution information (of
which four also have FIRST maps) are the only ones for which the
best guess can be made. These are described, individually, in what
follows:


\begin{description}
\item[\bf J0117+321:] In JVAS (e.g.\ Wilkinson et al.\ \cite{Wilkinson98}), this source shows up as compact ($<0.2\arcsec$). However, it is a GPS source and can be ruled out as candidate since it is likely too weak at low frequencies.
\item[\bf J0823+391:] Also in FIRST (resolved; $\sim30\arcsec$ wide large symmetric object with two edge-brightened lobes; 62~mJy/beam), this had further VLA observations done by Lehar et al.\ (\cite{Lehar01}) which show one of the lobes resolved ($\sim1\arcsec$ in size), the other compact ($<0.7\arcsec$; 4~mJy/beam), as well as a central core ($<0.7\arcsec$; $<1$~mJy/beam). It might contain VLBI compact components,
but is possibly too weak to cause confusion; on this basis, we rule
it out.
\item[\bf J0825+393:] In FIRST, it is a bright (1106~mJy/beam) unresolved source; mapped with VLBI, it looks like a compact (size $<0.07\arcsec$) steep spectrum source (Dallacasa et al.\ \cite{Dallacasa02}). A definitly confusing candidate that must be kept.
\item[\bf J1013+493:] A JVAS compact source ($<0.1\arcsec$ --- e.g.\ Patnaik et al.\ \cite{Patnaik92}), it is actually a VLBI calibrator with a size $<0.02\arcsec$ (Beasley et al.\ \cite{Beasley02}). In FIRST it shows up as a bright (266 mJy/beam) unresolved source. A definitly confusing candidate that must be kept.
\item[\bf J1215+331A:] Also known as NGC4203, this source has a FIRST map available (slightly resolved; 6~mJy/beam). It very likely contains a central compact core ($<1\arcsec$) with an inverted spectrum, possibly due to free-free absorption (e.g.\ Falcke et al.\ \cite{Falcke00}; Ho \& Ulvestad \cite{Ho01}). It shows an inverted spectrum at high frequencies, most likely too weak at low frequencies to confuse our observations, so ruled out.
\item[\bf J2152+175:] A core-plus-one-sided-jet VLBI source (e.g. Fey \& Charlot \cite{Fey97}), this source extends to very large structures becoming a narrow angle tailed large radio galaxy (Rector \& Stoke \cite{Rector01}). Both compact ($<0.2\arcsec$) and extended components are also seen in a VLA-A 8.4~GHz map (e.g. Browne et al.\ \cite{Browne98}). Its spectrum has a `knee' at $\sim1$~GHz, possibly peaking at $\la1$~MHz: a typical core+halo spectrum. A definitly confusing candidate that must be kept.
\item[\bf J2154+174:] Slightly resolved ($<0.01\arcsec$ size) with the VLBI (Beasley et al.\ \cite{Beasley02}) it is a VLA-A 8.4~GHz compact source ($<0.1\arcsec$; Browne et al.\ \cite{Browne98}). Its spectrum has a `knee' at $\sim1$~GHz, possibly peaking at $\sim10$~MHz (core+halo). A definitly confusing candidate that must be kept.
\end{description}

As regards the remaining, to first order, the answer lies in
the VLSS data. Only roughly half (15) of the 29 candidates are
stronger than the respective main source, all lacking high
resolution maps for compactness determinations. The question is,
then, {\it how to proceed?} In what follows, we will use all
existing information we can in order to {\em guess} the
compactness of each.

Taking advantage of existent spectral information, we
decided to use the spectral index value between VLSS and NVSS
($\alpha_{74}^{1400}$) of each candidate, as compared to the
corresponding value of the main source (if existent), as
indicative of the likelihood of a given candidate confusing
the observations or not. If $\alpha_{74}^{1400}$ is {\em steeper} for
the candidate than for the main source, we take it as likely to be more
extended, less compact, and hence less probable of confusing our
observations. Such comparison could be done for six of these 15
candidates, all rejected as confusing candidates. What
about the remaining nine? One of them (with
$\alpha_{74}^{1400}=1.0$) has no further information, but we rule
it out due to the comparison with the $\alpha_{fit}\simeq0.7$
overall 232--1400~MHz spectrum of the corresponding main source
which, in addition, has a Giga-hertz Peaked Spectrum (GPS)
core-like component at 1.4--8.4~GHz. Four other candidates do
have proper spectral information and the direct comparison using
$\alpha_{fit}$ rules them all out; three
have power-law spectra while the corresponding main source has a
spectrum with a `knee', suggesting a halo+compact core source ---
so, in these cases, the candidates are not likely to confuse our
observations (scintillations should come from the `core' component
in the main source). The last one, however, peaks at $\sim100$~MHz, while
the corresponding main source has a halo+core spectrum; hence,
both sources might have compact components and we do not know
which one is the stronger scintillator at 111~MHz. So, for caution,
we keep it as confusion candidate.

What about the 14 candidates that are weaker than the main source?
Although not confusing our observations, they might make some
relevant contribution to the total scintillating flux density of
the main source. Chasing their possible compactness properties, we
use $\alpha_{74}^{1400}$ as above\footnote{ And, in one case, also spectral information.} to rule out all
but four candidates that must be kept in the group because of their flatter values. As a matter
of fact, two of these candidates have high
resolution maps available (see Table~1) confirming them with
compact VLBI components.

The VLSS analysis is not yet complete, however: what about
non-detections? Candidates in this situation must be ruled out {\em only if}
the corresponding main source was indeed detected.
Out of the 736 candidates surveyed by the VLSS,
683 (93\%) are thus
ruled out in three different situations: i) 277 (NVSS) candidates reside in the 19 VLSS `pointings'
that found no confusing candidates at all; ii) 356 candidates in the remaining 14 VLSS
`pointings' with a detected main source stronger than each of them; iii) 50 `control' candidates, with FIRST and/or spectra
information (six have both; see Table~1).

With the hope of making use of the extant FIRST/NVSS data for
the 309 candidates left that were not surveyed by VLSS\footnote{ One of such candidates (J0823+391) was actually ruled out before thanks to a VLA map --- see text above.}, we
tried to define and calibrate criteria for ruling out candidates.
For this, we used both the 50 `control' candidates not detected in the VLSS and the 21 {\em extra} candidates in Table~1 that actually have both FIRST and radio spectra information (Appendix~B) --- six of the 27 candidates in Table~1 are included in the 50 `control' candidates (N.D. in Column~(7) of Table~1).
This, however, was not possible,
since both a combined classification and a separate one failed the
calibration tests. Hence, since there is no strong
statistical basis to rule out (or not) a
candidate using FIRST and/or spectral information, being
conservative, we keep all 333
remaining candidates, regardless of
their extra information. In a final attempt to split this number
into highest/lowest probabilities of confusing our observations,
we used 1.4~GHz NVSS flux density information (c.f. Appendix~A) to
reason as follows: if a source is too weak, its spectrum would
have to be too steep to reach `confusion levels', i.e., to have a
comparable low frequency flux density to the respective main
source. This time, we must set an arbitrary limiting value:
$\alpha=1.5$. Steeper candidates are rejected. Using, then, the
lowest frequency ({\it lo}) with measured flux density for each
respective main source (on 151--356~MHz), 262 candidates do not
reach 20\% of that value keeping $\alpha_{lo}^{NVSS}<1.5$ and are,
thus, rejected. The 71 candidates left can be further split into
16 included in Table~2, with the highest probability of causing
confusion (they reach main source flux densities within
$\alpha_{lo}^{NVSS}<1.5$), and 53 other (in the fields of 14 main
sources with ``confused?'' or ``confused'' in Column~(10) of
Table~3) which reach 20--100\% of each low frequency main source
flux densities within $\alpha_{lo}^{NVSS}<1.5$.

In Table~2 we present the final list of 23 high probability
confusing candidates, corresponding to the 11 main sources
signaled with ``confused'' in Column~(10) of Table~3. Thus
only about 23\% of our observed sources might have a good
probability of being confused. It is impossible to make any better
statement based on extant data since, to some degree, all 48
sources might be confused. Only detailed VLBI observations of {\em
all} 1046 candidates might establish definitive conclusions.


\section{Results}

\subsection{Overall}

The results of our observations are presented in Table~3. We
observed 50 sources from the sample of 55 sources in Augusto et
al. (1998) but only got scintillating flux density data for
48 (87\% completeness): five sources have not been observed
because of their high declinations ($\delta>70^{\circ}$),
resulting in too poor elongations\footnote{Ideally, these should be
on $22^{\circ}$-- $40^{\circ}$ at 111~MHz. Several other sources
at lower declinations had poor elongations. They did make it into
Table~3 since at least a scintillating flux
density ($S_c$) upper limit was possible to estimate for them. In the
strongest cases, a direct estimation of $S_c$ was
possible and, in even fewer, the actual total flux density ($S_t$)
was determined --- column (10) of Table~3.} --- B0205+722,
B0352+825, B0817+710, B0916+718, B1241+735; two other (B0905+420
and B1003+174) were confused by nearby strong VLBI sources
(B0904+417 and B1004+178, respectively), so no information about
$S_c$ is available for these either.

The IPS method requires knowledge of the upper limit for the (at
least; ideally the actual) size of the scintillating component of
a radio source in order to measure its flux density accurately.
We should gather as much size information as
possible for all compact components of each (e.g. Artyukh et al.
\cite{Artyukh99}). This was not
possible for the 17 sources in Table~3 (35\% of the total),
indicated with a star ($\star$) after their names, for which
either VLBI data are not available or there is still ambiguity in
identifying the scintillating component: their $S_c$ should have
their current uncertainties much reduced if/when those data are
collected. For example, the source B1058+245 has three
components Augusto et al. \cite{Augusto98}. The two at northeast have angular sizes
0.062$\times$0.022\arcsec (A) and 0.103$\times$0.053\arcsec (B),
and are separated by 0.047\arcsec. The southwest component has
size 0.314$\times$0.094\arcsec (C), and is 0.8\arcsec away from
the other two. Scintillations from {\em all} components will
add simultaneously, combining their flux densities. Hence,
we do not know which component(s) contributes most at
111~MHz, because we do not have spectral information for each
component. If it is component A, we get S$_c<$0.27Jy; if
component C we get S$_c<$0.5Jy. Thus, we put
the value $<$0.5Jy (hoping to improve it in the future)
in column (7) of Table~3.

When possible, we have thoroughly investigated the structure
of each source from the published (high frequency) VLBI-maps
(columns (4) and (5) in Table~3). We checked (when
possible) whether the spectra of compact components are peaked at
high frequencies. These components should not dominate at
low frequencies and we excluded them from further
consideration. We also excluded components which have
less than five times the flux density of any other
component. Among the remaining compact components, we tried to
find those with a comparatively high flux density and steep
spectrum at high frequencies and assumed that they have
power-law spectra down to low-frequency. Such an analysis allows
us to reveal one or several components of known angular size
dominating at 111~MHz.

In Column~(8) of Table~3 we show the total flux densities at
74~MHz from the VLSS while in Column~(9) we present the
$\alpha_{74}^{1400}$ spectral index with the help of the NVSS. Out
of the 48 sources, 33 (69\%) have, at least, some indication of
flux density at 74~MHz (seven are below $5\sigma$), while only one
(B0529+013) is not detected. The remaining 14 sources are not in
the current VLSS sky coverage.

\subsection{Case study: B0821+394}

We have chosen B0821+394 as a case study because it demonstrates
all features typical of scintillating sources. It has the
strongest scintillation in our sample (column (7), Table~3),
allowing us to even estimate $\sigma_{scint}$ `by eye' from
Fig.~1. It has enough total flux density to subtract the
background. Finally, it has a lot of observations at high angular
resolution, and therefore we can do an accurate analysis of its
structure in order to guess which compact components will dominate
at 111~MHz.
\begin{figure}[h]
\resizebox{\hsize}{!}{\includegraphics{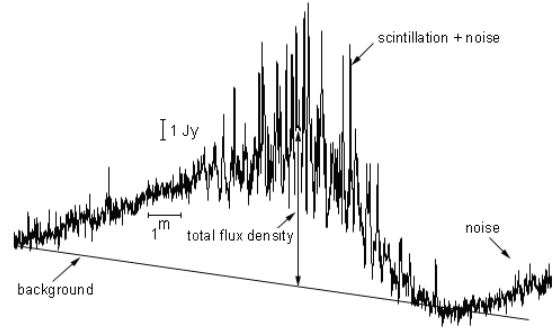}}
\caption{Primary record of the strong scintillating source
B0821+394. The comparison between ``scintillation+noise" and
``noise" gives us the possibility of estimating pure
scintillations. The details of reduction of such observations are,
for example, in Artyukh \& Tyul'bashev (\cite{Artyukh96}). }
\label{Fig1}
\end{figure}

Our observations of B0821+394 were obtained during six days at
elongations from $34^{\circ}$ to $46^{\circ}$ accumulating to a
total of 105 minutes. The value of $\sigma_{scint}$ varied
substantially from session to session of observations, therefore
we have a large error (25\%) in our estimation of $S_c$
--- Table~3. With an even larger error we could measure its
total flux density ($S_t$).

\begin{figure}[h]
\resizebox{\hsize}{!}{\includegraphics{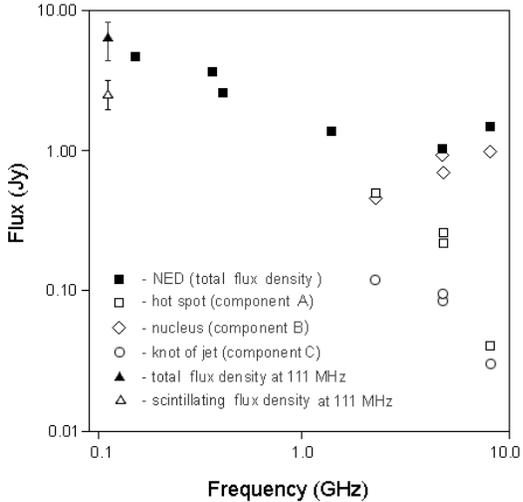}} \caption{
Spectrum of the source B0821+394 and spectra of its compact
components (NED --- NASA Extragalactic Database).} \label{Fig2}
\end{figure}
The radio source B0821+394 is a complicated SE-NW
core-plus-one-sided-jet with redshift 1.216 (Wills \& Wills
\cite{Wills76}; Augusto et al. 1998). This source has previously
been observed with high angular resolution (1.6~GHz --- MERLIN,
8.4~GHz --- VLA-A, 5~GHz --- VSOP; references in column~(5)
of Table~3) and from these data we can model it, to first order,
with three main components: A) NW `hot spot' with angular size
$11\times 9$~milliarcseconds (mas) at p.a.$=-38^{\circ}$ and
$\sim 250$ mas away from the nucleus; B) SE nucleus with size
$<0.3 \; \times<0.5$ mas; C) SE `knot' (start of jet) with size $2
\; \times<0.7$ mas at p.a.$=-18^{\circ}$, and $\sim 13$ mas away
from the nucleus. Since components A, B, and C are so close and compact,
B0821+394 will scintillate strongly from all,
simultaneously. Hence, the flux densities of these compact
components add to give the result in column~(7) of Table~3.
However, as we see next, only one of these components can dominate
at 111~MHz. Building the spectra of components A, B, and C
from information in the literature (Fig.2), we see that the
nucleus (B) has a GHz-peaked spectrum (decreasing to low
frequencies), while components A and C have power-law spectra. We
have an overall flat spectrum source, as was previously known
(Augusto et al. 1998). Our estimation of the total flux
density ($6.5\pm2$~Jy) agrees with other data while the
scintillating flux density from components A and C is
$2.5\pm0.6$~Jy.

\section{Discussion}

In a sense, this paper is a proof-of-concept for the application
of the IPS technique to flat-spectrum radio sources, which are
expected to cause rather more difficulty in the estimation of
total flux densities ($S_t$) at 111~MHz using such a method
since they are, generally, much weaker (in S$_t$) than
the steep-spectrum radio sources investigated earlier
(Tyul'bashev\& Chernikov \cite{Tyulbashev01}). We found
that one-fourth (13 out of 48) of the flat-spectrum sources
observed by us got estimates in $S_c$ that, {\em in the
worst scenario}, have errors smaller than 35\% (Table~3), with one
exception at $<50\%$. We were even able
to determine $S_t$ within 60\% errors for the five strongest sources.
How in the future can we improve our estimations of $S_c$? Simply
by getting proper multi-frequency high resolution observations
that might enable us to identify the scintillating
component(s) at 111~MHz. This next step is well under way (Augusto
et al., in prep.). As regards the 35 sources with upper limits
(only), we might improve them substantially or, better, transform
them into actual $S_c$ estimates, with the advent of high
resolution data.

The previous results, however, had to be strengthened,
due to the large LPA beam ($1\degr\times0.5\degr$), by making sure
that most targets were not affected by confusion. Due to the lack
of more appropriate surveys, 1046 confusion candidates were
identified by an extensive search in NVSS (surveying 100\% of
candidates), VLSS (70\%), FIRST (36\%), and NED. Of these, only
seven (0.7\%) have published high resolution maps (VLBI/VLA). It
is tantalizing that 97\% of the candidates residing in the VLSS
74~MHz surveyed areas (683 candidates, or 93\% of the total
number) were not detected ($S_{74}<0.5$~Jy/beam), while the
respective targets were so, all but one; and out of the remaining
3\% (29 candidates), using the steepness of $\alpha_{74}^{1400}$
(and VLBI maps for two) when compared with the corresponding main
source, only five (17\%) were not ruled out as causing
confusion. Four other candidates were maintained thanks to detailed
VLBI maps. Using detailed spectral information (for two) and VLA
maps, three further candidates were ruled out. Finally, 262 extra
candidates that cannot reach 20\% of the main source low frequency
({\it lo}) flux density within $\alpha_{lo}^{NVSS}<1.5$ were ruled
out; 53 that reach 20--100\% are low probability confusion
candidates while the remaining 16 join seven others from the
map/spectra selection (Table~2) as the highest probability
candidates for causing confusion: only 11 (23\%) of
the 48 main sources are thus affected.

As was pointed out in Augusto et al. (1998), 31 out of their 55
sources (56\%) have no data below $\sim$300~MHz. The observations
presented in this paper might be a breakthrough for establishing
the low-frequency spectra of the 55 sources in Augusto et al.
(1998), since 48 were observed, meaning 87\% completeness.
Relevant new information from our data comes from the
estimates on $S_c$ at 111~MHz as compared with the
total flux densities at 74/151~MHz from the literature
(e.g. Augusto et al.\ 1998 and Table~3) --- we can place an
approximate upper limit on the flux densities of extended
low surface brightness components, for the sources\footnote{Using
the minimum possible value as a lower limit for the
flux density in compact components; e.g.\ $0.7\pm0.2$~Jy gives us
a lower limit of 0.5~Jy.} B0116+319 ($\la0.4$~Jy),
B0824+355 ($\la1.5$~Jy), and
B1211+334 ($\la1.5$~Jy).

Knowledge of the low-frequency end of the radio spectrum of a
radio source (and its components) is vital before fitting any
synchrotron emission model to gain knowledge about its physical
properties. Our objective, in due course, is to make such fits for
all 55 sources. There is potential for all but one
source since, in addition to the 48 presented in this paper, six
out of the seven left out actually have 151~MHz total flux
densities in the literature. Since these exist in two
main types (core+(distorted)-jets; compact/medium symmetric
objects --- believed to be the precursors of large FRI/FRII radio
galaxies), we think we can contribute to clarifying the origin of
this subset of active galactic nuclei, at least as regards their
emission mechanisms.

\acknowledgements{ We aknowledge an anonymous referee for
helpful suggestions and comments. We are grateful to
R.D.Dagkesamanskii for attention to our work and for hosting
P.\/Augusto during his visit. S.A.\/Tyul'bashev acknowledges
the programs ``Solar Wind" and ``Extended sources" from the
Russian Academy of Sciences for the partial support of this work.
P.\/Augusto acknowledges the research grant PESO/P/PRO/15133/1999
from the Funda\c c\~ao para a Ci\^encia e a Tecnologia (Portugal).
This research made use of the United States Naval Observatory
(USNO) Radio Reference Frame Image Database (RRFID) and of the
NASA Extragalactic Database (NED). }

\onecolumn
\appendix

\section{The flux densities of the confusion candidates}
As regards NVSS flux densities, out of the 1046 candidates,
950 (91\%) have 1.4~GHz flux densities $<37$~mJy. This leaves 96
strong ($\geq37$~mJy) candidates\footnote{The division between `strong' and `weak' candidates is quite arbritary. The value of 37~mJy was picked simply because above it the flux density distribution is more discrete, with some holes, while below it all unit values have (at least) one source and, obviously, the further down the scale the more the sources.},
of which 28 are also in the
VLSS. It is tantalizing that, out of these 96 `strong' candidates,
only four, in three
pointings, have NVSS 1.4~GHz flux densities stronger than the
corresponding main source with flux density ratios 1.6, 1.1, 2.9
and 1.3, respectively.

For all three FIRST types [i) unresolved ($U$); ii) slightly
resolved ($SR$); iii) resolved ($R$)] --- see Appendix~B, the
weakest detected source has 1~mJy/beam.
In Figure~3 we present the flux density distributions for each type.
Comparing the flux density distributions up to
S$_{1.4}=15$~mJy/beam, we exclude 29\% of unresolved sources, 14\%
of slightly resolved ones, and 19\% of resolved sources. Going
further down to S$_{1.4}\leq4$~mJy/beam, the respective exclusion
rates are 65\%, 45\%, and 39\%, thus showing a trend for the
weakest sources to be resolved.

The comparison of the VLSS flux densities between
the candidates and each corresponding main source is only possible
for 13 pointings (out of 19), corresponding to 20 candidates (out
of 29), since for the remaining there are no VLSS flux density
measurements of the main source. Overall, the flux density ratios
are in the range 0.2--4.1 with all but two candidates in the interval 0.4--2.6.
Hence, the typical candidate-main source flux density ratio is
within a factor of about 2.5.

\section{FIRST/spectral classification}

As regards to the use of spectral information, in the hope of
applying a similar spectral criterion to the one applied for the
29 VLSS candidates (Sect.~3), as before, depending on the number
and range of the data points, we split the 60 candidates with
spectral information into two large groups (usually the main
source has more data points than each corresponding candidate and
includes data at all available frequencies): {\it two data points
--- group I} --- two-frequency spectral index calculation,
compared with the same spectral index for the main source; {\it
three to five data points---group II} --- a linear regression is
made (a global spectral index is fitted) and the result is
compared with the one obtained by applying the same technique to
the corresponding main source, using the same frequency range.
Then, depending at which frequencies they have data, we split them
further into the following seven subgroups (between brackets the
number of candidates inside each subgroup) : {\it Ia)}
1.4--2.7~GHz (1); {\it Ib)} 1.4--4.85~GHz (8); {\it Ic)} (318 or
365 or 408) to 1400~MHz (27); {\it Id)} 151--1400~MHz (8); {\it
IIa)} 3-point fit; 151--408~MHz to 1.4--4.85~GHz (12); {\it IIb)}
4-point fit; 74--365~MHz to 4.85--8.4~GHz (3); {\it IIc)} 5-point
fit; 151~MHz to 4.85~GHz (1). The 60 candidates were then
classified as ``steep'' or ``flat'' {\em relative} to the
respective main source (c.f.\ Table~1). ``Steep'' cases would be
expected to be ruled out as candidates, while ``flat'' ones would
be kept in.

In order to use the FIRST information, we analysed the ``postage
stamps'' available from the Internet (not contour plots), and
decided to split the morphologies of the 271 candidates found into
three groups: i) unresolved sources ($<$5\arcsec in size; $U$) ---
89 candidates (33\%); ii) slightly resolved sources
($\sim$5\arcsec in size; $SR$) --- 107 candidates (39\%); iii)
resolved sources ($>$5\arcsec in size; $R$) --- 75 candidates
(28\%) (Fig.3).
\begin{figure}[h]
\setlength{\unitlength}{1cm}
\begin{picture}(16.0,10.0)
\put(-1.3,11.1) {\includegraphics{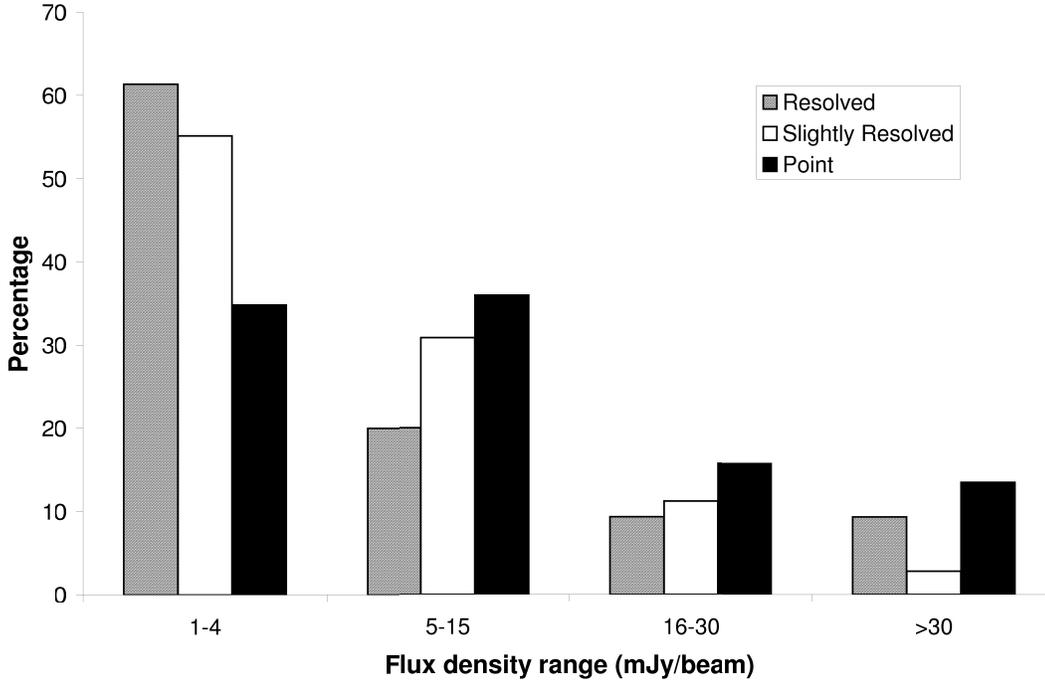}}
\end{picture}
\caption{The FIRST flux density distributions of the 271
candidates with such data, divided into the classifications
`resolved', `slightly resolved', and `point'. The averages are,
respectively, $11$~mJy/beam, $9$~mJy/beam, and $20$~mJy/beam.}
\label{Fig3}
\end{figure}

 Noting that FIRST alone gives a very poor
indication on the existence or not of VLBI compact structure, we
decided to include also information from NVSS, using the ratio of
both 1.4~GHz flux densities ($S_{FIRST}/S_{NVSS}$) to define a
{\it compactness} ($c$) parameter. Since the NVSS beam is nine
times the width of the FIRST beam (81 times in area) it is also a
lot more sensitive to extended structure. Thus, we would expect a
point source to have $c=1$ while a very resolved source would have
$c\ll1$. Indeed, although with large dispersions, the averages are
$c=0.8$ ($U$), $c=0.7$ ($SR$), and $c=0.4$ ($R$). We have, then,
decided to use these criteria together in order to define {\em
resolved} candidates (to be ruled out as confusing candidates) as
the ones with\footnote{For the rest of this Appendix we used the logical symbols $\vee$ (for OR) and $\wedge$ (for AND) in order to compactify the exposition.} $(R \vee SR) \wedge c\leq0.6$ and {\em unresolved}
(to be kept in) the ones with $(U \vee SR) \wedge c\geq0.9$. Any
other situations would not be considered, since they were too
ambiguous. It must be emphasized that even a $U \wedge c=1.0$
candidate is not guaranteed to have compact VLBI structure, since
the FIRST resolution is 5\arcsec. Variability complicates the
picture: some sources have been observed some years apart between
the two surveys. For example, values of $c>1$ (27 candidates in
271; 10\%) must be due to variability. We expect FIRST unresolved
sources (likely containing cores) to be more variable than
extended ones; indeed, 15 of the 27 variable candidates (56\%) are
`unresolved' while `slightly resolved' are the remaining (44\%)
--- there are no variable `resolved' candidates.

The vital move then was, by using cross-information for the
71 `control' candidates (including the ones in Table~1) as
calibration, to test our criteria for deciding on a candidate
status. Starting with the seven candidates (Table~1) that have
high resolution maps available: four with detailed spectral
information (e.g. `inverted spectrum') had correct decisions made,
finding compact components there; four with FIRST information also
reach consistency: $R \wedge c=0.3$ for the large, resolved source
and $(SR \vee U) \wedge c=0.9$--1.0 for the other three, with VLBI
components. Unfortunately, it was also evident some inconsistency
in two of the latter for which a steep spectrum corresponds to $U
\wedge c=0.9$--1.0; clearly, the spectrum of some candidates might
be too complicated (with too many components, eventually including
compact ones) to conclude anything just from such analysis. This
is further stressed if we look at the remaining 20 candidates of
Table~1: while nine are immediately ignored as FIRST ambiguous,
only other nine of the remaining 11 have consistent
FIRST/spectral information ($(SR \vee R) \wedge c\leq0.6$ and
steep; $(SR \vee U) \wedge c\geq0.9$ and flat). Taking our
`calibration' further, we also looked, separately as it could only
be, at FIRST and spectral decisions for the remaining 44 `control'
candidates: i) 13 candidates with FIRST data split into two
variable, four ambiguous, two with $(SR \vee U) \wedge c\geq0.9$
and five with $(SR \vee R) \wedge c\leq0.6$ ; ii) 31 candidates
with spectral data split into 19 steep, three flat ($Ic$); two
steep, one flat ($II$); and six flat ($Ib$), of which four have
inverted spectra. To add even more information on this, we have
used 95 candidates with FIRST $(SR \vee R) \wedge c\leq0.6$ and
$(SR \vee U) \wedge c\geq0.9$ classifications (which are not
mentioned anywhere else in this paper) which split into two
suspiciously size-comparable samples, with 50 in the former
classification and 45 in the latter. Hence, our FIRST and/or
spectra criteria fail too often to be of any use for decisions on
candidates which only have these data available, in addition to
NVSS.

\clearpage \onecolumn

\tablecaption{The 27 confusion candidates that have both radio
structure and radio spectra information. {\bf (1)}: The J2000.0
name of the candidate; marked with an asterisk are `control'
candidates (see main text); {\bf (2):} 2000.0 right ascension from
the NVSS; {\bf (3):} 2000.0 declination from the NVSS; {\bf (4):}
short description of the source morphology with high resolution
maps (VLBI and VLA), including sizes; CSS: compact steep spectrum
source; {\bf (5):} morphological description from FIRST, if
surveyed (LSO: large symmetric object; U: unresolved; SR: slightly
resolved; R: resolved); `bright' means $S_{1.4}^{FIRST}>170$~mJy
--- all other sources have $S_{1.4}^{FIRST}<110$~mJy; {\bf (6):}
the compactness parameter ($S_{1.4}^{FIRST}/S_{1.4}^{NVSS}$); {\bf
(7):} the spectral index as compared with the main source value
calculated from the VLSS (if surveyed; N.D. means no detection in
VLSS) and the NVSS; {\bf (8):} compared spectral indices from
other frequencies (subgroups as in Appendix~B); when detailed
spectral information exists, it is described; GPS --- Gigahertz
Peaked Spectrum Source.}
\tablehead{\hline \multicolumn{1}{c}{\bf
(1)} & {\bf (2)} & {\bf (3)} & {\bf (4)} & {\bf (5)} & {\bf (6)} &
{\bf (7)} & {\bf (8)}
\\
\multicolumn{1}{c}{Candidate} & R.A. & Dec & high resolution maps
& FIRST & $c$ & $\alpha_{74}^{1400}$ & other spectral information
\\ \hline}
\tabletail{\hline}
\begin{supertabular}{cccccccl}
J0117+321$^*$ & 01 17 55.13& +32 06 22.9 &
\multicolumn{1}{l}{VLA-A compact ($<0.2\arcsec$)} & --- & --- &
N.D. & GPS \\ J0737+238 & 07 37 52.12 & +23 52 44.9 &---& R&
0.2& ---& {\it IIa} steep \\ J0822+078 & 08 22 06.81 & +07 53
46.1 &---& U& 0.8& ---& {\it Ic} steep \\ J0822+079A & 08 22
50.00 & +07 58 30.4& ---& R& 0.4& ---& {\it Ib} steep \\
J0823+391 & 08 23 23.96 & +39 06 29.7 &\multicolumn{1}{l}{three
$\sim1$~mJy comps.;} & LSO (R) & 0.3 & ---& {\it IIb} steep \\
& & & \multicolumn{1}{l}{two compact ($<0.7\arcsec$)} & & & & \\
J0823+082 & 08 23 39.14 & +08 14 30.8 &---& SR& 0.8& ---& {\it
IIa} steep \\
J0825+393 & 08 25 23.64 & +39 19 45.6
&\multicolumn{1}{l}{CSS ($<0.07\arcsec$)} & bright (U) & 0.9 &
---& {\it IIa} steep \\
J0827+390 & 08 27 14.52 & +39 05 46.1 &--- & R & 0.2& ---& {\it
Ic} steep \\ J0828+354 & 08 28 47.97 & +35 24 26.1 &---& SR&
0.8& ---& {\it Id} steep \\
J0836+554 & 08 36 20.38 & +55 28
58.6 &---& bright (U)& 0.8& ---& {\it IIa} steep; main peaks \\
& & & & & & & at $\sim100$~MHz \\
J0837+557 & 08 37 52.99 & +55 45 43.9 &---& R& 0.4& ---& {\it Id}
steep \\
J1012+287 & 10 12 06.73 & +28 42 43.0 &---& U & 0.9&
---& {\it IIa} flat \\
J1013+493 & 10 13 29.97& +49 18 40.8 &\multicolumn{1}{l}{VLBI cal
($<0.02\arcsec$)} & bright (U) & 1.0&---& {\it IIb} steep \\
J1212+330B$^*$ & 12 12 53.20& +33 01 23.8& --- & U& 0.3& N.D. &
{\it Id} steep \\
J1215+331A$^*$ & 12 15 05.23& +33 11 52.7
&\multicolumn{1}{l}{compact, inverted} & SR& 1.0& N.D.& {\it Ia}
flat --- inverted \\
& & & \multicolumn{1}{l}{spectrum core ($<1\arcsec$)} & & & & spectrum \\
J1233+536A & 12 33 11.38& +53 39 56.7 &---& U& 0.7& ---& {\it Id}
steep \\
J1233+536B & 12 33 41.89& +53 37 23.8 &---& R & 0.6&
---& {\it IIa} steep \\
J1235+538 & 12 35 13.48& +53 49 06.8 &---& R& 0.6& ---& {\it Id}
steep \\
J1236+534 & 12 36 34.22& +53 25 41.5 &---& R& 0.4&
---& {\it Id} steep \\
J1237+535 & 12 37 50.35& +53 33 38.2 &---& R& 0.5& ---& {\it IIb}
steep \\
J1238+534 & 12 38 08.16& +53 25 56.0 &---& U& 0.4&
---& {\it Ib} flat \\
J1318+197A$^*$ & 13 18 20.01 & +19 46 47.1 &---& U& 0.4& N.D.&
{\it Id} steep \\ J1342+340$^*$ & 13 42 46.48& +34 02 22.9
&---& U & 0.9& N.D.& {\it Id} steep \\ J1629+212 & 16 29
47.56& +21 17 17.7 &---& bright (SR)& 0.7& steep & convex: peaks
at $\sim10$~MHz; \\
& & & & & & & main at $\sim100$~MHz \\
J1721+561$^*$ & 17 21 49.68& +56 07 50.1 &---& U& 1.0& N.D.& {\it
IIb} steep \\
J2152+175 & 21 52 24.81& +17 34 38.2
&\multicolumn{1}{l}{narrow angle tailed} & ---& ---& flat &
halo+core spectrum; main \\
& & & \multicolumn{1}{l}{VLBI radio galaxy} & & & & with similar spectrum \\
J2154+174 & 21 54 40.83& +17 27 49.6
&\multicolumn{1}{l}{VLBI-size ($<0.01\arcsec$)} & ---& ---& flat
& halo+core spectrum; main \\
& & & & & & & with similar spectrum \\
\hline
\end{supertabular}

\vspace{1cm} \clearpage
\tablecaption{The 23 sources that most likely confuse our
observations. {\bf (1):} The J2000.0 name of the candidate; the
sources marked with an asterisk are also listed in Table~1; {\bf
(2):} 2000.0 right ascension from the NVSS; {\bf (3):} 2000.0
declination from the NVSS; {\bf (4):} The VLSS 74~MHz flux
density; {\bf (5):} The NVSS 1.4~GHz flux density; {\bf (6):}
short description of the reason for keeping the candidate as a
confusing source; CSS: compact steep spectrum source.}
\tablehead{ \hline {\bf (1)} & {\bf (2)} &{\bf (3)} &{\bf (4)}
&{\bf (5)} & \multicolumn{1}{c}{\bf (6)} \\ J2000.0 name & R.A. &
Dec & S$_{74}$ & S$_{1400}$ & \multicolumn{1}{c}{Reason for
keeping in}
\\ & & & (Jy) & (mJy) &
\\ \hline}
\tabletail{\hline}
\begin{supertabular}{cccccp{6cm}}
J0046+318 & 00 46 40.93 & +31 51 25.2 & 0.87 & 195 & peak $\sim100$~MHz vs.\ halo+core \\
J0639+357 & 06 39 29.80& +35 43 36.8 &--- & 62 & S$_{1400}$ extrapolation ($\alpha<1.5$ test) \\
J0639+355 & 06 39 58.31 & +35 32 56.5 &--- & 94 & S$_{1400}$ extrapolation ($\alpha<1.5$ test) \\
J0642+355 & 06 42 43.21 & +35 33 01.1 &--- & 66 & S$_{1400}$ extrapolation ($\alpha<1.5$ test) \\
J0643+354 & 06 43 48.68 & +35 28 34.0 &--- & 63 & S$_{1400}$ extrapolation ($\alpha<1.5$ test) \\
J0822+078$^*$ & 08 22 06.81 & +07 53 46.1 & --- & 72 & S$_{1400}$ extrapolation ($\alpha<1.5$ test) \\
J0822+079A$^*$ & 08 22 50.00 & +07 58 30.4& --- & 118 & S$_{1400}$ extrapolation ($\alpha<1.5$ test) \\
J0823+082$^*$ & 08 23 39.14 & +08 14 30.8 &--- & 138 & S$_{1400}$ extrapolation ($\alpha<1.5$ test) \\
J0825+393$^*$ & 08 25 23.64 & +39 19 45.6 &--- & 1198 &VLBI map: CSS\\
J0836+554$^*$ & 08 36 20.38 & +55 28 58.6 & --- & 288& S$_{1400}$ extrapolation ($\alpha<1.5$ test) \\
J1012+287$^*$ & 10 12 06.73 & +28 42 43.0 &--- & 70 & S$_{1400}$ extrapolation ($\alpha<1.5$ test) \\
J1013+493$^*$ & 10 13 29.97& +49 18 40.8 &--- & 265 &VLBI map: calibrator \\
J1233+536A$^*$ & 12 33 11.38& +53 39 56.7 & --- & 63 & S$_{1400}$ extrapolation ($\alpha<1.5$ test) \\
J1233+536B$^*$ & 12 33 41.89& +53 37 23.8 &--- & 81 & S$_{1400}$ extrapolation ($\alpha<1.5$ test) \\
J1236+534$^*$ & 12 36 34.22& +53 25 41.5 &--- & 136 & S$_{1400}$ extrapolation ($\alpha<1.5$ test) \\
J1237+534 & 12 37 02.14& +53 25 28.2 &--- & 72 & S$_{1400}$ extrapolation ($\alpha<1.5$ test) \\
J1237+535$^*$ & 12 37 50.35& +53 33 38.2 &--- & 208 & S$_{1400}$ extrapolation ($\alpha<1.5$ test) \\
J1238+534$^*$ & 12 38 08.16& +53 25 56.0 &--- & 115 & S$_{1400}$ extrapolation ($\alpha<1.5$ test) \\
J1802+034 & 18 02 51.09& +03 27 02.9 & --- & 178 & S$_{1400}$ extrapolation ($\alpha<1.5$ test) \\
J1859+630 & 18 59 48.89& +63 04 36.5 &1.31 & 168 &$\alpha_{74}^{1400}$ criterion (flatter than main source) \\
J2151+177 & 21 51 45.06& +17 43 07.2 &0.43 & 42 &$\alpha_{74}^{1400}$ criterion (flatter than main source) \\
J2152+175$^*$ & 21 52 24.81& +17 34 38.2 & 1.56 & 680 &VLBI map: compact components + $\alpha_{74}^{1400}$ criterion (flatter than main source) \\
J2154+174$^*$ & 21 54 40.83& +17 27 49.6 &1.38 & 294 &VLBI map: compact components $\alpha_{74}^{1400}$ criterion (flatter than main source) \\
\hline
\end{supertabular}
\clearpage
\tablecaption{The scintillating flux
density (or upper limit) of the scintillating component(s) at
111~MHz for 48 of the 55 sources in Augusto et al. (1998). {\bf
(1):} B1950.0 and J2000.0 names; when a star ($\star$) follows, it
means that the source has the potential to get improved values of
flux densities limited/measured, when relevant VLBI data are
available. {\bf (2):} The amount of individual records. {\bf (3):}
The elongation range during the observations. {\bf (4):} The
maximum size of the scintillating component, estimated from the
high-resolution information on the references listed in {\bf (5)}
and (still) unpublished VLBI maps (Augusto et al., in prep.). {\bf
(5):} References for the high resolution maps used, with code
numbers translated at the footnote of this table. {\bf (6):} 111
MHz dispersion ($\sigma_{scint}$) during a scintillation across
the {\em full} observation range (c.f. Fig.~1). {\bf (7):} 111 MHz
scintillation flux density ($S_c$) measurements with error, or
upper limit. {\bf (8):} Total flux density at 74~MHz (from the
VLSS); sources with a range given are detected but not above
$5\sigma$. {\bf (9):} Spectral index between 74~MHz (VLSS) and
1.4~GHz (NVSS). {\bf (10):} General comments/information where we
give: i) the total 111 MHz flux densities ($S_t$) for the five
sources for which this was possible to measure (SNR$>30$); ii)
CSO-MSO (compact-medium symmetric object) classification, after
Augusto et al. (1998) and Augusto et al. (1999); iii) other
information. } \tablehead{\hline \multicolumn{1}{c}{\bf (1)} &
{\bf (2)} & {\bf (3)} & {\bf (4)} & {\bf (5)} & {\bf (6)} & {\bf
(7)} & {\bf (8)} & {\bf (9)} & {\bf (10)} \\
\multicolumn{1}{c}{Names} & $N$ & $\epsilon$ & $\theta$ &
References & $\sigma_{scint}$ &$S_c$ & S$_{74}$ &
$\alpha_{74}^{1400}$ &Comments \\ & & ($^{\circ}$) & ($''$) &
&(Jy) &(Jy) & (Jy)& & \\ \hline} \tabletail{\hline}
\begin{scriptsize}
\begin{supertabular}{lccccccccc}
B0046+316/J0048+319& 7 &41 &$<0.2 $ & 1,2,7,15 &$<0.15$ &
$<0.22$ & 0.1--0.5 & & CSO \hspace{5mm} confused \\
B0112+518/J0115+531& 5 &61 &$<0.2 $ & 1,2 & 0.22 & $0.7\pm0.2$ & 1.52 & 0.42 & MSO? \\
B0116+319/J0119+321$^{\star}$ & 9 &23 &$<0.2 $ &1,2,3,5,14,18 & 0.6 & $0.75\pm0.15$ & 1.06 & $-0.31$ & CSO; $S_t=1.7\pm1$~Jy\\
B0127+145/J0129+147$^{\star}$ & 5 &40 &$<0.2 $ & 1,2 &$<0.25$ & $<0.5$ & 4.73&0.62 & \\
B0218+357/J0221+359& 13 &22--51 &$<0.01 $ & 2,3,6,9,10,13,16 &$<0.25$ & $<0.3$ & 3.57 &0.25 & \\
B0225+187/J0227+190& 12 &32--50 &$<0.03 $ & 1,2 &$<0.15$ & $<0.21$ & 0.1--0.5 & & CSO or MSO \\
B0233+434/J0237+437& 9 &40--54 &$<0.01 $ & 1 &$<0.2$ & $<0.33$ & 0.1--0.5 & & CSO \\
B0345+085/J0348+087$^{\star}$ & 18 &19--23 &$<0.2 $ & 1,2 &$<0.25$ & $<0.47$ & 0.65 & 0.34 & \\
B0351+390/J0355+391& 10 &25--40 &$<0.1 $ & 1,2 &$<0.25$ & $<0.37$ & 0.66 & 0.41 & \\
B0418+148/J0420+149$^{\star}$ & 10 &28--33 &$<0.2 $ & 1,2 &$<0.25$ & $<0.49$ & 1.30 & 0.32 & \\
B0429+174/J0431+175& 5 &25 &$<0.1 $ & 1,2 &$<0.5$ & $<0.75$ & 0.1--0.5 & & \\
B0529+013/J0532+013$^{\star}$ & 18 &22--53 &$<0.2 $ & 1,2 &$<0.4$ & $<0.7$ & $<0.1$ & & confused? \\
B0638+357/J0641+356& 8 &22--29 &$<0.1 $ & 1,2 & 0.25 &$0.3\pm0.1$ &--- & & MSO? \hspace{5mm} confused \\
B0732+237/J0735+236& 16 &15--38 &$<0.1 $ & 1,2 &$<0.17$ & $<0.24$ & --- & & CSO \hspace{5mm} confused? \\
B0819+082/J0822+080& 6 &25--52 &$<0.1 $ & 1,2 &$<0.3$ & $<0.55$ & ---& & CSO or MSO \hspace{5mm} confused \\
B0821+394/J0824+392& 6 &34--46 &$<0.01 $ & 2,3,4,8,10 &1.3 &$2.5\pm0.6$& ---& & $S_t=6.5\pm2$~Jy \hspace{5mm} confused \\
B0824+355/J0827+354$^{\star}$ & 9 &25--42 &$<0.1 $ & 1,2,11 &0.6 &$0.7\pm0.2$ & --- & & MSO \hspace{5mm} confused? \\
B0831+557/J0834+555& 8 &40--50 &$<0.01 $ &2,3,5,9,19,20 &0.75 &$1.26\pm0.25$ & ---& & $S_t=11.3\pm2.5$~Jy \hspace{5mm} confused \\
B1010+287/J1013+284& 8 &19--59 &$<0.02 $ & 1 &$<0.15$ & $<0.2$ & ---& & CSO \hspace{5mm} confused \\
B1011+496/J1015+494$^{\star}$ & 10 &36--75 &$<1 $ & 1,2 &$<0.15$ & $<0.75$ & --- & & confused \\
B1058+245/J1101+242$^{\star}$ & 8 &19--29 &$<0.3 $ & 1,2 &$<0.23$ & $<0.5$ & 1.24 &0.34 & MSO? \\
B1143+446/J1145+443$^{\star}$ & 12 &29--84 &$<0.1 $ & 1,2 &$<0.1$ & $<0.3$ & --- & & confused? \\
B1150+095/J1153+092$^{\star}$ & 7 &26 &$<0.1 $ & 1,2,17 &0.28 & $0.34\pm0.1$ & ---& & confused? \\
B1211+334/J1214+331& 23 &32--74 &$<0.05 $ &1,2 &0.25 &$0.42\pm0.08$ & 2.07 & 0.13 & \\
B1212+177/J1215+175& 7 &30 &$<0.05 $ &1,2 &0.39 &$0.55\pm0.11$ & 1.87 &0.21 & CSO \\
B1233+539/J1235+536& 15 &51--82 &$<0.1 $ & 1 &$<0.15$ & $<0.37$ & ---& & CSO or MSO \hspace{5mm} confused \\
B1317+199/J1319+196& 15 &28--52 &$<0.1 $ &1,2 &0.23 &$0.31\pm0.06$ & 2.04& 0.35 & \\
B1342+341/J1344+339& 15 &43--60 &$<0.03 $ & 1,2 &$<0.1$ & $<0.22$ & 0.1--0.5 & & \\
B1504+105/J1507+103& 21 &31--53 &$<0.1 $ & 1,2 &$<0.2$ & $<0.36$ & ---& & CSO \hspace{5mm} confused? \\
B1628+216/J1630+215$^{\star}$ & 10 &42--56 &$<0.2 $ & 1,7 &$<0.15$ & $<0.31$ & 1.67 & 0.40 & MSO? \\
B1638+124/J1640+123& 13 &34--57 &$<0.05 $ & 1,2 &0.6 &$0.8\pm0.4$ & 2.24 &0.03 & \\
B1642+054/J1644+053$^{\star}$ & 16 &28--60 &$<0.2 $ & 1,2 &$<0.2$ & $<0.34$ & ---& & \hspace{5mm} confused? \\
B1722+562/J1722+561& 16 &78--84 &$<0.1 $ & 1 &$<0.15$ & $<0.75$ & 1.01& 0.55 & \\
B1744+260/J1746+260& 25 &60 &$<0.1 $ & 1,2 &$<0.15$ & $<0.3$ & 0.83& 0.29 & \\
B1801+036/J1803+036& 9 &36 &$<0.03 $ & 1,2 &$<0.15$ & $<0.21$ & ---& & MSO? \hspace{5mm} confused \\
B1812+412/J1814+412& 6 &67 &$<0.1 $ & 1,2,11 &0.6 &$1.7\pm0.8$ & 2.97& 0.48 & $S_t=10\pm5$~Jy \\
B1857+630/J1857+630& 7 &84 &$<0.1 $ & 1 &$<0.25$ & $<1.2$ & 2.47& 0.72 & confused \\
B1928+681/J1928+682& 13 &81--88 &$<0.1 $ & 1,2 &$<0.2$ & $<0.8$ & 1.09& 0.22 & CSO \\
B1947+677/J1947+678$^{\star}$ & 7 &85 &$<0.2 $ & 12 &$<0.2$ & $<1$ &0.1--0.5 & & MSO? \\
B2101+664/J2102+666& 8 &76--87 &$<0.03$ & 1 &$<0.15$ & $<0.4$ &0.1--0.5 & & \\
B2112+312/J2114+315& 7 &46 &$<0.1 $ & 1,2 &$<0.2$ & $<0.4$ & 1.80& 0.51 & \\
B2150+124/J2153+126$^{\star}$ & 6 &38 &$<0.2 $ & 1,2 &$<0.3$ & $<0.7$ & 2.29& 0.57 & \\
B2151+174/J2153+176$^{\star}$ & 3 &30 &$<0.2 $ & 1,2 &$<0.4$ & $<0.7$ & 3.27& 0.90 & confused \\
B2201+044/J2204+046& 7 &40 &$<0.1 $ & 1,2 &$<0.5$ & $<0.9$ & 2.68& 0.59 & \\
B2205+389/J2207+392& 16 &47--63 &$<0.1 $ & 1,2 &$<0.4$ & $<0.8$ & 1.57& 0.35 & \\
B2210+085/J2213+087& 6 &30 &$<0.1 $ & 1,2 &$<0.25$ & $<0.36$ & 0.87 & 0.41 & \\
B2247+140/J2250+143$^{\star}$ & 6 &20 &$<0.2 $ & 1,2 &1.3 &$2.2\pm0.3$ & 4.81&0.30 & $S_t=5.5\pm2$~Jy \\
B2345+113/J2347+115$^{\star}$ & 15 &35--60 &$<0.2 $ & 1,2 &$<0.3$ & $<0.7$ & 0.93&0.34 & CSO or MSO \\
\end{supertabular}
\end{scriptsize}
\begin{table}[b]
\begin{tabular}{l}
1. Augusto et al. \cite{Augusto98} \\
2. JVAS, CLASS surveys
(Patnaik et al. \cite{Patnaik92}, Browne et al. \cite{Browne98},
Wilkinson et al. \cite{Wilkinson98}, Myers et al. \cite{Myers03})
\\
3. Fomalont et al. \cite{Fomalont2000} \\
4. Fey \& Charlot
\cite{Fey2000} \\

5. Fey \& Charlot \cite{Fey97} \\

6. Kellermann et al. \cite{Kellermann98} \\

7. ftp://rorf.usno.navy.mil/RRFID/index.html \\

8. Thakkar et al. \cite{Thakkar95} \\

9. Polatidis et al. \cite{Polatidis95} \\

10. Xu et al. \cite{Xu95} \\

11. Henstock et al. \cite{Henstock95} \\

12. Sykes \cite{Sykes97}\\

13. Patnaik et al. \cite{Patnaik93} \\

14. Wrobel \& Simon \cite{Wrobel86} \\

15. Unger et al. \cite{Unger84} \\

16. Kemball et al. \cite{Kemball01} \\

17. Morabito et al. \cite{Morabito86} \\

18. Altschuler et al. \cite{Altschuler95} \\

19. Pearson \& Readhead \cite{Pearson88} \\

20. Faison \& Goss \cite{Faison01} \\


\end{tabular}
\end{table}
\end{document}